\documentstyle[mprocl]{article}

\input BoxedEPS
\SetRokickiEPSFSpecial  
\HideDisplacementBoxes

\def\be{\begin{equation}}
\def\ee{\end{equation}}
\def\bea{\begin{eqnarray}}
\def\eea{\end{eqnarray}}

\let\hat\widehat 
\let\bar\overline 
\def\hhhh{ \hat{\bar{\psi}}\,\hat{\bar{\psi}}\, \hat{\psi}\, \hat{\psi}}
\def\pppp{ \bar{\psi} \,\bar{\psi}\, \psi\, \psi}
\def\bp{ \bar{\psi}\, \psi}
\def\sq{\mbox{\scriptsize$\sqcap$\llap{$\sqcup$}}}

\def\DD{{\cal D}}
\def\RT{\mathop{\rm Re \, Tr}}

\def\notA{/ \llap{$A$}}
\def\notpt{/ \llap{$\partial$}}

\begin{document}

\title{HADRON STRUCTURE IN LATTICE QCD: \\ Exploring the Gluon Wave
Functional}

\author{John W. Negele}

\address{Center for Theoretical Physics, \\Laboratory
        for Nuclear Science, and Department of Physics, \\
        Massachusetts Institute of Technology, Cambridge, 
        Massachusetts 02139 U.S.A.\\E-mail: negele@mitlns.mit.edu\quad MIT
CTP\# 2999}


\maketitle\abstracts{ 
The use of lattice QCD to understand hadron structure is described, with
particular emphasis on exploring the role of glue.
}


\section{Introduction}

This conference on  excited states of the
nucleon emphasizes the wealth of information about hadron structure 
accessible empirically through clever experiments and frontier
electromagnetic probes. For a theorist, it also highlights the concomitant need
for understanding the remarkably rich structure of hadrons from first
principles.

Lattice QCD is the only known way to solve, rather than model, QCD\null. Given
the  development of the  tools of lattice field
theory and advances in computer technology that now make Terascale
lattice computation affordable, the time is  right for a major  
initiative in hadronic physics. Thus, it is particularly noteworthy for
participants at this conference at Jefferson Lab that Jefferson Lab, MIT, and
the Lattice Hadron Physics Collaboration, representing twenty-two theorists
at thirteen institutions, have undertaken such an initiative and proposed 
dedicated Lattice QCD  clusters with a total capacity of 480 Gflops to study
hadron structure. The understanding and support of the experimental
community for such an initiative is crucial, and I hope to provide  in this talk
glimpses of some of the physics that will be possible. 

One of the most striking features of the quark and gluon structure of the
nucleon is the extremely important role of gluons as essential dynamical
degrees of freedom. In most familiar many-body problems, the bosons whose
exchange mediates the interaction between fermions may be subsumed into
an effective two-body interaction and hence do not appear directly as 
essential degrees of freedom. In atomic physics, photon exchange is
subsumed into a static Coulomb potential and one solves the Schr\"odinger
equation in the presence of Coulomb forces. Similarly in nuclear physics, to a
good approximation nuclear structure may be calculated as a many-body
problem in the presence of   a strong, state-dependent but static
nucleon-nucleon potential.  In both cases, the mass of an atom or nucleus
comes overwhelmingly from the masses of the fermionic constituents. 

Nucleons are strikingly different.  We know experimentally that
approximately half of the momentum and angular momentum of a nucleon is
carried by its gluons. Furthermore, the mass of a nucleon would hardly be
altered if the masses of its quarks decreased from their physical scale of the
order of a few MeV to zero, so the mass arises almost completely from the
gluonic interactions between fermions. Hence, although we normally only
think of the fermionic component of the atomic or nuclear wave function, in
the case of the nucleon it is essential to understand the gluon component of
the wave function. As I will describe below, lattice QCD provides an excellent
framework to explore this physics.

In lattice QCD, an observable is evaluated by defining quark and gluon
variables on the sites and links of a space-time lattice, writing a Euclidean path
integral of the generic form 
\begin{eqnarray}
 \langle Te^{-B\hat{H}} \hhhh \rangle 
&=&Z^{-1} \int  \DD(U) \DD (\bp) e^{-\bar{\psi} M(U) \psi - S(U)} \pppp
\label{E:JN:1}  \nonumber\\
 &=& Z^{-1} \int  \DD(U) e^{\ln \det M(U) - S(U)} M^{-1}
(U) M^{-1} (U)
\nonumber
\end{eqnarray}
and evaluating the final integral over gluon link variables $U$ using the
Monte Carlo method.  The link variable is $U=e^{iagA_\mu(x)}$, the Wilson
gluon action is $S(U)=
\frac{2n}{g^2} \sum_{\sq} (1-\frac{1}{N} \RT U_{\sq})$ where
$U_{\sq}$ denotes the product of link variables around a single
plaquette, and $M (U)$ denotes the discrete Wilson approximation to the
inverse propagator such that in the continuum limit,  $M (U) \to m+ \notpt  +
\ ig \, \notA$.  For those unfamiliar with lattice QCD, an elementary
introduction at the level required for this work is presented in
ref.\cite{Negele:1997bi}  For our present purposes, it is useful to note
that evolution in Euclidean time provides a convenient filter to extract the
ground state from an arbitrary state of specified quantum numbers, since \\
$e^{-\beta H} |\Psi\rangle = \sum_n e^{-\beta E_n} c_n \psi_n \to
\mbox{const} 
\times \psi_0 + {\cal O}(e^{-\beta(E_1-E_0)})$.

\section{Lattice Wave Functions}

 The wave function plays a central role 
in our understanding of many-body systems.  Familiar examples in 
nonrelativistic quantum mechanics include the
shell model for nuclear and atomic 
systems, the BCS wave function for superconductivity and the phases
of liquid He, and the Laughlin wave function for the quantum Hall
effect.
It
is thus natural to seek analogous understanding of hadronic structure
in terms of quark and gluon wave functionals, which become wave functions
when defined on a discrete lattice.

In nonrelativistic quantum mechanics, we define an N-particle wave function
as the overlap between a state in which the positions of N particles are
specified and the state $|\psi \rangle$, $ \langle x_1, \ldots, x_N|\psi \rangle
= \psi(x_1, \ldots, x_N)$. When it is impractical to deal with the exact wave
function, one may alternatively study a variational wave function specified
by some set of variational parameters $\{\alpha_1, \ldots, \alpha_n\}$ and
evaluate the inner product $\psi_{\mathrm trial}(\alpha_1, \ldots, \alpha_n)|\psi
\rangle$.  The ``best" approximation is obtained by varying the $\alpha$'s to
maximize the overlap. A familiar nuclear physics example is approximating a
light nucleus by a harmonic oscillator wave function, and varying the size
parameter to obtain the maximum overlap.

For pure Yang-Mills gauge theory on a lattice, we specify the gauge field
$A_{\mu}(x_n)$ for each space-time direction $\mu$ and site $x_n$, so the
lattice  wave function is the overlap $\langle A_0(x_1), \ldots,  
A_4(x_N)| \psi \rangle$ corresponding to the wave functional
$\psi[A_{\mu}(x)]$ in the continuum limit. In QCD, it is convenient
to use occupation number representation for the  quarks, and indicate the
lattice sites that contain quarks or antiquarks, so that the wave function is
written: 
$\langle q(x_1), \ldots,   q(x_n), \bar q(y_1), \ldots,   q(y_m), A_0(x_1),
\ldots,\linebreak    A_4(x_N)\, | \, \psi \rangle$.  
Whereas it will be straightforward to examine
components of the wave function involving a small number of quarks and
antiquarks at specified positions, it will not be practical to keep track of all
the components of the gauge field $A_{\mu}(x_n)$ (which exceed 10$^5$ on a
small 16$^4$ lattice).  Hence, in general, we will need to consider some simple
or physical {\it Ansatz} for $A_{\mu}(x_n)$, and I will give illustrative
examples below. 

\begin{figure}[t!]
\begin{center}
\BoxedEPSF{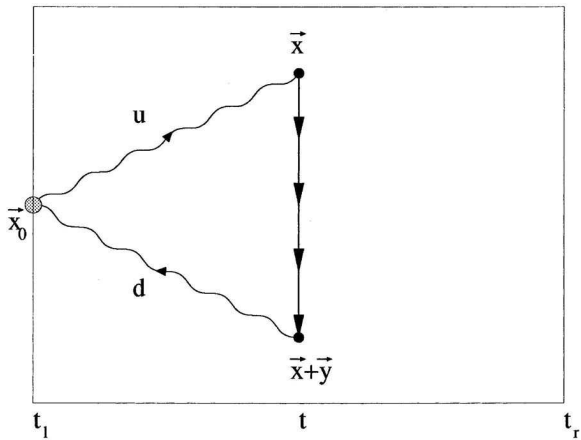 scaled 1000}
\end{center}
\caption{Lattice calculation of the meson wave function, $\psi_s$,
corresponding to the overlap between the meson ground state on the left
with a configuration containing a $\bar q$-$q$ pair connected by a string of
glue on the right. 
\label{fig:1}}
\end{figure}

Given the form of a trial wave function   
$$\bigl|q(x_1), \ldots,   q(x_n), \bar q(y_1), \ldots,   q(y_m), A_0(x_1), \ldots,  
A_4(x_N)\bigr\rangle$$  the overlap with the ground state hadron wave
functions is calculated on the lattice as sketched schematically in
Fig.~\ref{fig:1}. A source with the quantum numbers of the desired meson is
placed at the leftmost time 
$t_l$, and propagation to time $t$ near the middle of the lattice filters out
the ground state. On time slice $t$, one constructs an operator corresponding
to the trial wave function. In the case considered in Fig.~\ref{fig:1},  a quark
and antiquark are separated by distance $\vec y$, the cm variable $\vec x$
is integrated to project onto zero momentum,  and the gluon wave function is
the ground state vacuum configuration of gluons multiplied by a string of
flux created by a Polyakov line as shown. Summing over an ensemble of
gluon configurations then yields the quantum mechanical result for the
overlap. 

To elucidate the role of the gluon component of the hadron wave functional, it
is instructive to display the results for several alternative approximations to
the gluon field. To highlight the basic ideas, I will show some pedagogical
results calculated previously in the meson sector by Kien Boon
Teo\thinspace\cite{Teo:1994zu,teo}. With the current interest in nucleon
structure and the resources presently available, analogous calculations should
certainly be carried out for nucleons.  In particular, it is instructive to
consider the following three definitions of variational wave functions.

\subsection{Axial gauge or string wave function $\psi_s$}

The conventional gauge-invariant wave function, which corresponds to 
the Bethe-Salpeter amplitude in the axial gauge $A_y=0$, 
is given by
\begin{equation}
 \psi_s(y) = \bigl\langle0\bigl| \bar{q}(0)\Gamma e^{\int^{y}_{0} dz A_y(z)}
 q(y) \bigr|h\bigr\rangle 
\end{equation}
where $\Gamma$ depends on the Dirac structure of the ground state hadron
$|h\rangle$. The exponential corresponds to a
 product of U links as shown in Fig.~\ref{fig:1}  and
the implied gluonic component of $\psi_s$ is that of an infinitely thin
string of glue. The square of the resulting wave function specifies the
probability that a $q \bar q $ pair is separated by a distance $y$ {\it and}
that the glue forms a narrow string connecting them. 

\subsection{Coulomb gauge-fixed amplitude $\psi_c$}

Gauge fixing to Coulomb gauge surrounds each quark with the gluons
corresponding to the static Coulomb field.
In the Abelian case we can write down the following explicit
expression:
\begin{equation}
  \psi_c(y) = \bigl\langle0\bigl| \bar{q}(0)\Gamma 
 e^{\int d^3z \vec{E}_{\mathrm
static}(\vec{z}).\vec{A}(\vec{z})}q(y)\bigr|h\bigr\rangle
\end{equation}
where $\vec{E}_{\mathrm static}$ is the static Coulomb field of an $e^+e^-$
pair separated by distance~$y$. Since 
$$\int \vec A \cdot \vec E_{\mathrm static} = \int
\vec A \cdot  \nabla \phi = -\int  (\nabla \cdot \vec A) \phi $$  
it is evident that the exponent vanishes in Coulomb gauge.

\begin{figure}[htb]
\begin{center}
\BoxedEPSF{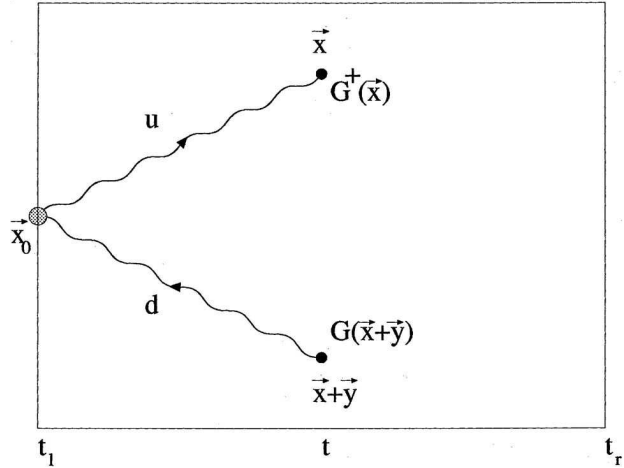 scaled 1000}
\end{center}
\caption{Lattice calculation of the meson wave function, $\psi_c$,
corresponding to the overlap between the meson ground state (\emph{left})
with a vacuum gluon configuration in Coulomb gauge (\emph{right}). 
\label{fig:2}}
\end{figure}

 For QCD, the implied gluonic component of $\psi_c$ is obtained by the lattice
measurement shown in Fig.~\ref{fig:2} with the gluon field at time slice $t$
fixed to Coulomb Gauge.  As discussed later, the Coulomb gauge result has a
well-behaved continuum limit. 


\subsection{Adiabatic Wave Function $\psi_a$}

A more physical definition of a gluon wave function is
the adiabatic wave function obtained by letting the 
gluonic component be the ground state distribution of gluons in the
presence of a $q\bar{q}$ pair at separation y:
\begin{equation}
\psi_a(y) = \lim_{n\rightarrow \infty} {\bigl\langle0\bigl|\bar{q}(0) \Gamma
             S_y(t,n)q(y)\bigr|h\bigr\rangle \over
  W(y,2n)^{1\over 2} }  
\end{equation}
where
$S_y(t,n)$ = $U_{0}^{(0,t\rightarrow 0,t+n)}\cdot U_y^{(0,t+n\rightarrow
y,t+n)}\cdot U_{0}^{+(y,t\rightarrow y,t+n)} $
 and $W(y,2n)$ is a Wilson
loop of size $y \times 2n$. When the temporal links are extended (by
increasing~$n$) far enough, the gluons in the vacuum adjust themselves to the
presence of the $q\bar{q}$ pair, forming the ground state of a flux tube
connecting two fixed color charges. 
\begin{figure}[tbh]
\begin{center}
\BoxedEPSF{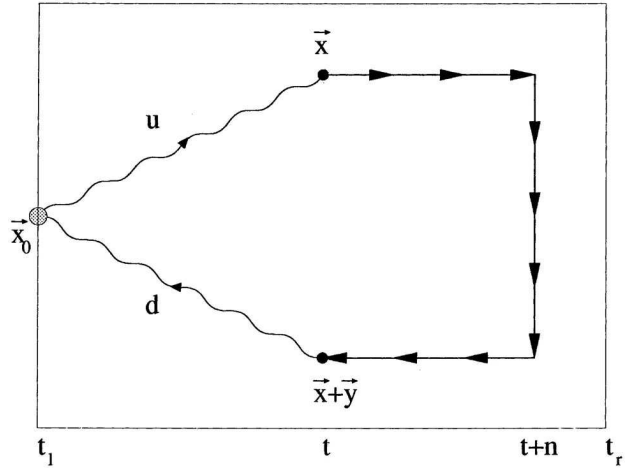 scaled 1000}
\end{center}
\caption{Lattice calculation of the adiabatic meson wave function, $\psi_a$,
corresponding to the overlap between the meson ground state on the left
with a configuration containing a static $\bar q q$ pair connected by a
ground state flux tube on the right. 
\label{fig:3}}
\end{figure}

\section{Density-Density Correlation Functions}

\begin{figure}[ht]
\begin{center}
\BoxedEPSF{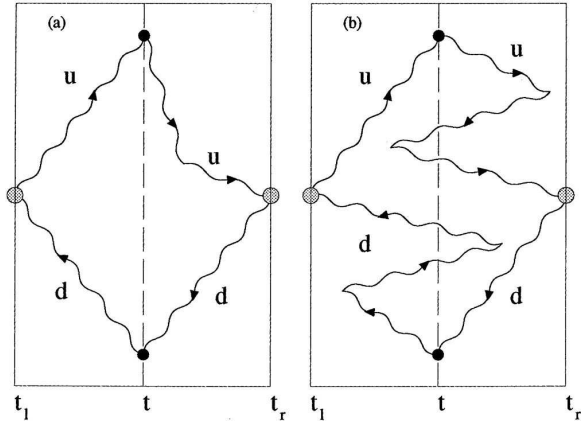 scaled 900}
\end{center}
\caption{Two representative time histories contributing to the meson
density-density correlation function. Diagram (a) represents a
contribution from the $\bar q q$ sector whereas (b) corresponds to the
$\bar q \bar q \bar q q q q$ sector.  
\label{fig:4}}
\end{figure}

Since the density-density correlation function is a gauge-invariant
quantity that describes the distribution
of quarks in a hadron and reduces to the square of the
 wave function in nonrelativistic quantum mechanics, we have for comparison
also calculated
\begin{equation}
 \langle\rho(x)\rho(0)\rangle =
\bigl\langle
h\bigl|\bar{u}\gamma_{0}u(x)\bar{d}\gamma_{0}d(0)\bigr|h\bigr\rangle
\end{equation}
and its projection
 $\langle\rho\rho\rangle_{q\bar{q}}$
onto
the $q\bar{q}$ subspace. Computationally the projection is performed
by using hard-wall boundary conditions to exclude 
those paths in which the propagators can cross
the time slice~$t$ at which $\langle\rho\rho\rangle$
is evaluated.  Figure~\ref{fig:4} shows
representative time histories corresponding to the density-density
correlation function in the $\bar q q$ sector, diagram (a), where only one
quark and antiquark propagator cross time slice $t$, and a general
contribution, diagram (b), containing multiple quark-antiquark excitations at
time~$t$.

\section{Lattice Calculation of Meson Wave Functions\\ and Correlation
Functions}

\begin{figure}[ht]
\begin{center}
\BoxedEPSF{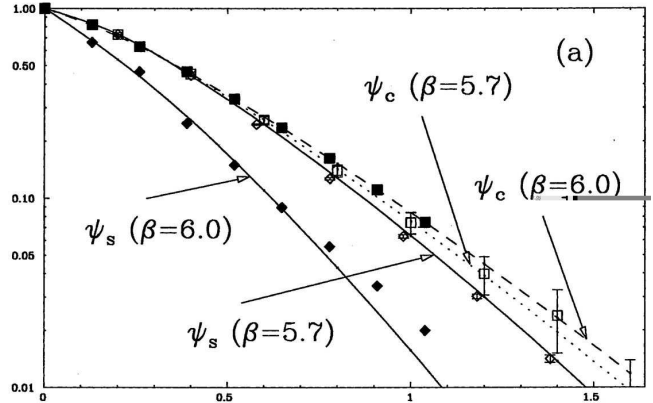 scaled 900}
\end{center}
\caption{Pion $\psi_s$ and $\psi_c$ for $\beta=5.7$
(Teo and Negele\thinspace\protect\cite{Teo:1994zu}) and 
$\beta=6.0$ 
 (extracted from Hecht and DeGrand\thinspace\protect\cite{Hecht}) versus
$q\bar{q}$ 
separation y. The lines are to guide the eye.  \label{fig:5}}
\end{figure}

\begin{figure}[ht]
\begin{center}
\BoxedEPSF{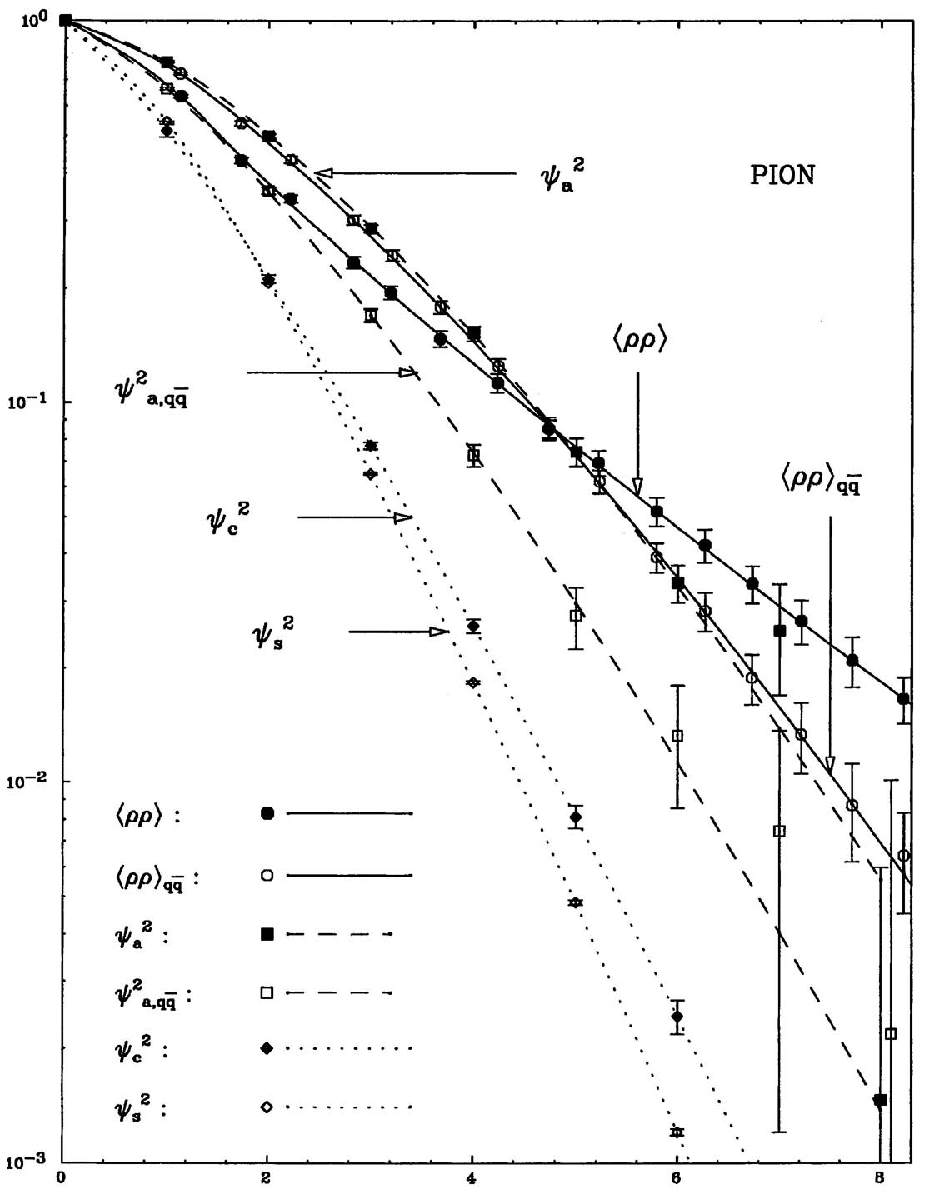 scaled 1000}
\end{center}
\caption{Comparison of  $|\psi|^2$ and density-density correlation functions
 for  the pion as a function of the
$q\bar q$ separation in units of
0.2 fm.  Lines connecting Monte Carlo results are to guide the
eye. \label{fig:6}}
\end{figure}

\begin{figure}[ht]
\begin{center}
\BoxedEPSF{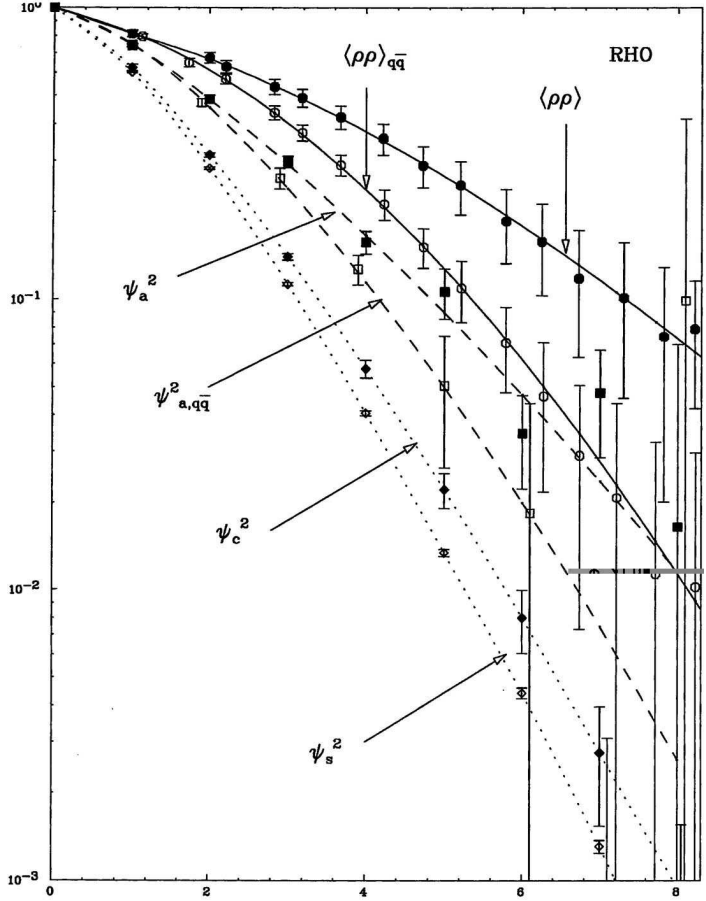 scaled 1000}
\end{center}
\caption{Comparison of  $|\psi|^2$ and density-density correlation functions
 for  the rho meson as a function of the
$q\bar q$ separation in units of
0.2 fm.  Lines connecting Monte Carlo results are to guide the
eye.  \label{fig:7}}
\end{figure}

Figures ~\ref{fig:5}--\ref{fig:7}  show the results of calculating the three
wave functions and two density-density correlation functions described
above in quenched lattice QCD\thinspace\cite{Teo:1994zu,teo}.  

All propagators were calculated with twenty quenched configurations at
$\beta=5.7$ on a $16^3\times16$ lattice using the Wilson action for both
gluons and quarks and  quark masses of 40, 95, and 170 MeV. We
computed $\psi_s$ using a string of links and explicitly fixing to
Coulomb gauge for $\psi_c$. $\Gamma=\gamma_5$ and
$\Gamma=\gamma_i$ are used for $\pi$ and $\rho$ mesons respectively.

Figure \ref{fig:5}
compares $\psi_s$ and $\psi_c$ calculated on the lattice for
two different values of $\beta$ with a  quark mass
170 MeV.
 The $\psi_c$'s agree to within
error bars, indicating a well-behaved continuum limit. In contrast, $\psi_s$
decreases with increasing~$\beta$ because of lattice artifacts
that diverge as the lattice spacing approaches zero. As discussed below, the
small overlap between a thin string of glue and the gluons in the ground state
hadron accounts for much of the falloff of $\psi_s$ with $q\bar{q}$
separation.

Figures \ref{fig:6} and  \ref{fig:7}
 show the results for the squares of wave functions and density-density
correlation functions for
$\pi$ and
$\rho$ mesons, respectively.  All quantities are normalized
 to 1 at zero
$q\bar{q}$ separation. We display results for a quark mass of 40 MeV,
where errors are in better control, instead of the extrapolated
chiral limit which differs inconsequentially.

The dramatic difference between $\psi_a^2$, $\psi_c^2$, and $\psi_s^2$
shows that the gluon flux tube of the adiabatic wave function
produces a much larger overlap with the
gluons in the ground state hadron than does the Coulomb or string
wave function. Beyond 1 fm, the adiabatic flux tube is favored by more
than an order of magnitude. Physically,  there is a substantial probability
to find a quark-antiquark pair separated by 1 fm. connected by a
physical adiabatic flux tube, but a very small probability to find them joined
by an unphysically narrow string of flux. Also
note, as mentioned in connection with Fig~\ref{fig:5}, that the string of flux
becomes increasingly unfavorable as one approaches the continuum limit.

 The role of multiple quark-antiquark excitations in the density-density
correlation function is evident from the fact that $\langle\rho\rho\rangle$ falls off
more slowly than
 $\langle\rho\rho\rangle_{q\bar{q}}$ at large distances. 
For both the pion and rho, $\psi_a^{2}$ is 
close to $\langle\rho\rho\rangle_{q\bar{q}}$, suggesting that the adiabatic flux tube is a
reasonable approximation to the full gluon wave function.

\section{ Overlap  with Variational Wave Functions}

A complementary means of studying the overlap of a trial gluon wave
function with a hadronic ground state is through the calculation of two-point
\begin{figure}[ht]
\begin{center}
\BoxedEPSF{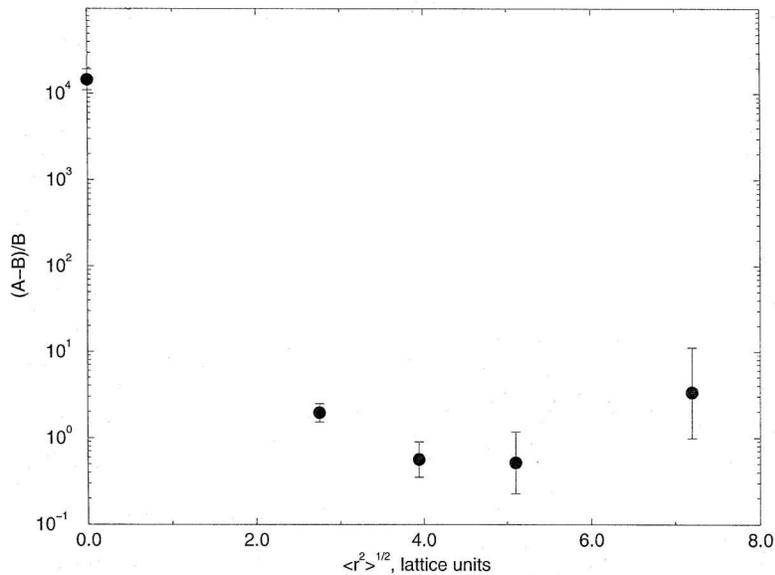 scaled 1000}
\end{center}
\caption{Ratio of excited state content to ground state content of
$\psi_{\mathrm trial}$ as a function of the size of the trial function in units of 0.1 fm. 
\label{fig:8}}
\end{figure}
correlation functions. Let $J^{\dagger}$ be a source with the desired quantum
numbers and denote $|\psi_{\mathrm trial} \rangle = J^{\dagger} |0\rangle$. For
example, $J^{\dagger}$ could be built from point sources, Gaussian
distributed sources in Coulomb gauge, or gauge-invariant Wuppertal-smeared
sources\thinspace\cite{gusken} of the form:
$$\psi(x) = \sum_{x'}\Bigl(1 + \alpha\sum_i\bigl[U_i(x)\delta_{x',x+i} +
U_i^{\dagger}(x-i)\delta_{x',x-i}\bigr]\Bigr)^N q(x')$$
where $q(x)$ is the quark creation operator.  The two-point function may be
written
$$ \langle J(T) J^{\dagger}(0) \rangle = c\sum_n\bigl| \bigl\langle
\psi_{\mathrm trial}\bigm| n
\bigr\rangle\bigr|^2 e^{-E_nT}.$$
At  $T=0$, the value of the correlation function is $A = c
\sum_n|\langle\psi_{\mathrm trial}|n\rangle|^2$. The large $T$ behavior of the
correlation function is
$c|\langle
\psi_{\mathrm trial}| 0 \rangle
|^2 e^{-E_0T}$, and extrapolating this behavior back to $T=0$ yields $B = c
|\langle\psi_{\mathrm trial}|0\rangle|^2$.  Hence one may study the overlap of
$\psi_{\mathrm trial}$ with the physical ground state by calculating ${B/ A} 
=  |\langle\psi_{\mathrm trial}|0\rangle|^2$ and optimize it with respect to the
variational parameters in $\psi_{\mathrm trial}$

Figure \ref{fig:8} shows the ratio of overlap with excited states to the
overlap with the ground state, $ ({A-B})/{B} =  {\sum_{n>0}
|\langle\psi_{\mathrm trial}|n\rangle|^2}\big/{|\langle\psi_{\mathrm
trial}|0\rangle|^2}$ for a nucleon created with Wuppertal smeared sources of
different rms radii\cite{Dolgov:1999js}. The striking fact here is that, whereas
for point sources  the excited state content is
$10^4$ larger than the ground state content, for the optimal smearing  the
overlap with the ground state, ${B/ A}  = 
|\langle\psi_{\mathrm trial}|0\rangle|^2$, is approximately 65\%. In other words, the
sum of the sets of gauge-field links generated by the smearing provides a
reasonable physical approximation to the behavior of the gauge fields in the
nucleon. This suggests that a systematic study of variational wave functions
has the potential to yield even larger overlaps and to provide a valuable tool
for exploring the nucleon's gluon wave function.

\section{Conclusions}

Lattice QCD provides a powerful tool for exploring hadron
structure and, in particular, the gluon wave functional. 
The exploratory results  for meson wave functions shown here
display large effects of the  gluonic components in
three definitions of  hadron wave functions, and the
adiabatic flux tube wave function yielded the most physical results.
Quantitative measurement of the overlap of a trial wave function with the
ground state may be obtained from two-point correlation functions,
providing the opportunity for quantitative study of variational wave
functions.

\section*{Acknowledgments}
This work is supported in part by funds provided by the U.S. Department of
Energy (DOE) under cooperative research agreement \#DF-FC02-94ER40818.

\vfill
\eject

\end{document}